\documentclass[preprint,
aps,showpacs,tightenlines,
nofootinbib,showkeys,byrevtex]{revtex4}

\usepackage{graphicx}%
\usepackage{dcolumn}
\usepackage{amsmath}
\usepackage{epsfig}
\usepackage{latexsym}

\def\darr#1{\raise1.5ex\hbox{$\leftrightarrow$}\mkern-16.5mu #1}
\def\){\right)} 
\def\({\left(} 
\def\]{\right]} 
\def\[{\left[}

\def\si{{}^1\kern-.14em S_0}
\def\siii{{}^3\kern-.14em S_1}
\def\diii{{}^3\kern-.14em D_1}

\def\pone{{}^3\kern-.14em P_1}
\def\pzero{{}^3\kern-.14em P_0}
\def\ptwo{{}^3\kern-.14em P_2}

\def\nrcpt{NR\raise.4ex\hbox{$\chi$}PT\ }

\def\ltap{\ \raise.3ex\hbox{$<$\kern-.75em\lower1ex\hbox{$\sim$}}\ }
\def\gtap{\ \raise.3ex\hbox{$>$\kern-.75em\lower1ex\hbox{$\sim$}}\ }

\newcommand{\beq}{\begin{eqnarray}}
\newcommand{\eeq}{\end{eqnarray}}

\begin{document}
\bibliographystyle{plain}
\preprint{LBL-50618}

\title{Dilute resonating gases and the third virial coefficient}

\author{Paulo F. Bedaque \footnote{Email: {\tt pfbedaque@lbl.gov}}
and
Gautam Rupak \footnote{Email: {\tt grupak@lbl.gov}} }

\affiliation{
Lawrence Berkeley National Laboratory, Berkeley, CA, U.S.A. 94720 \\
}


\begin{abstract}
We study dilute gases with short range
interactions and large two-body scattering lengths. At
temperatures between the condensation temperature and the scale set by the
range of the potential  there is a high degree of
universality. The first two terms in the expansion of thermodynamic functions 
in  powers of
the fugacity $z$, which  measures the diluteness of the system, are
determined by the scattering length only. The term proportional to
$z^3$ depends only on one new parameter describing the three-body
physics. We compute the third term of the expansion and show that,
for many values of this new parameter, the $z^3$ term
 may be the dominant one.

\end{abstract}

\pacs{ }
\keywords{ }
\maketitle

\begin{section}{Introduction}
\label{introduction}

Only on special circumstances  the
thermodynamic functions of a system can be evaluated starting from the microscopic
interactions. 
However, in many situations  only a few characteristics of the microscopic
interactions are relevant. All systems sharing these same
microscopic characteristics have then the same macroscopic behavior, that is,
there is a certain degree of universality. An example of such a
system is a gas of particles with short range interactions. As
long as the density and temperature are such that the typical
wavelength $\lambda$ of the particles is much larger than the
range $R$ of the forces, the details of the interaction potential
is largely irrelevant and the thermodynamics is determined by the
two-body scattering length $a$, up to corrections of order
$\lambda/R$. This observation has been used 
 both for
fermionic and bosonic gases 
since the $1950$'s ~\cite{bogoliubov_bosons,lee_yang_bosons,lee_huang_yang_bosons,wu_bosons,
brueckner_bosons,beliaev_bosons,lieb_bosons,hugenholtz_pines_bosons,braaten_hans_largea} in
the computation of dilute gas properties in an expansion in
powers of $n a^3$, where $n$ is the density of particles.

 Typically, the size of the two-body scattering length is comparable to the range of the force,
$a\approx R$. However, there are important situations where the
interactions are fine-tuned in such a way as to make $a\gg R$,
even though other low-energy parameters like the effective range
$r_0$, etc., still have the size expected on  dimensional grounds
$r_0\approx R$. We have in mind two of these situations. The first
one, a gas of neutral atoms, has received enormous experimental
and theoretical attention recently. The range of the interactions
 between the atoms is set by the length scale $R\sim (M C_6/\hbar^2)^{1/4}$ 
of the van der Wals
force $-C_6/r^6$. In some atomic species like $^{87}$Rb and $^4$He
the fine tuning needed for large values of $a$ is provided by Nature. 
In the case of  $^4$He 
atoms for instance, $a=104 \mathring{A}$, a value much larger than  $R\sim 5
\mathring{A}$ or the effective range $r_0\approx 7 \mathring{A}$. More importantly, an
external magnetic field can be applied in  order to artificially
modify the scattering length, making it a {\it tunable} parameter
(Feshbach resonance) . The second example of unnaturally large
scattering lengths is  a gas of  neutrons (and protons, if the
Coulomb force can be disregarded). The range of the nuclear forces
is of the order of the Compton wavelength of the pion ($R\sim
\hbar/m_\pi\sim 1.5 fm$), but the scattering lengths are significantly
larger ($5.42$ fm for the proton-neutron in the spin-triplet
state to $-18.8$ fm  for neutron-neutron in the spin-singlet state).

 The typical momentum of the particles in a gas is set by the larger of the
 inverse interparticle distance
$n^{1/3}$ and the inverse thermal wavelength $\lambdabar
=(M T/2\pi)^{-1/2}$ (from now on we will use $\hbar=1$). 
In the natural case, $a\approx R$,
 the
universal regime occurs for $1/\lambdabar, n^{1/3} \alt
1/a\approx 1/R$. This regime is essentially perturbative in the sense
that, at any given order of the expansion in powers of $1/R$, a
properly set up diagrammatic expansion involves only a finite
number of diagrams ~\cite{hans_dilute}.
On the other hand when $1/\lambdabar, n^{1/3} \approx 1/R$ the details of 
the particle 
interactions are important and each system should be studied in a
case by case basis. The presence of large scattering lengths opens
up an intermediate regime, where the typical wavelength
$\lambda$ of the particles is comparable to $a$, but still smaller
than $R$ . The problem is no longer perturbative, as evidenced by
the fact that bound states of size $\approx a$ are expected, but some
degree of universality should still hold. This regime can be
attained at  very low temperatures and  
a moderately high densities $ n a^3\sim 1 \agt n
R^3$  or at  a moderately high temperatures 
$ a/\lambdabar \sim 1
\agt R/\lambdabar$ and a very low densities. It is an outstanding problem to
understand the first of these regimes, since, most likely,
many-body correlations are not suppressed, and a number of  publications
have appeared recently on the subject ~\cite{giorgini_etal,kalos_etal,holland_largea_fermi, 
holland_largea_bosontofermi,pandharipande_largea,braaten_nieto_largea}. In this paper we will
consider the second case, namely a dilute ($n a^3\alt 1$),
moderately hot ($(M T/2\pi)^{1/2}\ a\approx 1$) gas of resonating
$a\agt R$ particles.

Since we are considering low densities, it is convenient to  phrase our discussion
in terms of the virial expansion, where the different
thermodynamic functions are expressed as power series on the
fugacity $z=e^{\beta \mu}$, where $T=1/\beta $ is the temperature
and $\mu$ the chemical potential. The particle density $n$ and the pressure $P$ are given by
\beq n&=&\frac{1}{\lambdabar^3}(b_1 z+ 2 b_2 z^2 + 3 b_3 z^3 + \cdots)
\nonumber\ ,\\
P&=& \frac{T}{\lambdabar^3}(b_1 z+ b_2 z^2 + b_3 z^3 + \cdots)
\label{virialexpansion}\ .\eeq

The usefulness of the virial expansion resides on the fact that,
for small densities, $n\alt \lambdabar^{-3}$, $z$ is also small $z\ll
1$. The coefficients $b_l$ contains contributions coming from
$m$-particle correlations for all $m\leq l$ so its computation
involves the solution of the $l$-body problem. The
calculations of $b_1$ and $b_2$ are rather easy and can be done
analytically in the system considered here. The computation of
$b_3$ is a little more involved: it  includes some numerical
integrations, and it is related to the quite unusual properties
of the three resonating particle system.

There is a general formula relating $b_3$ to S-matrix elements ~\cite{dashenetal_virial}. 
However, the relevant matrix elements 
relevant here are the ones describing a variety of processes like three-to-three-particle scattering, 
dimer break up in a collision 
to a particle, etc.. It seems to us that the equivalent method followed below implicitly 
includes all these processes in a simple way. Also, since it is an straightforward 
application of standard
effective  field theory and many-body physics methods, it may have some methodological interest.

We will use the language of effective field theory, which is 
natural when exploring universal, low energy, large distance
properties that are independent of the short distance details. We will
work at the leading order in the  small momentum expansion where
the Lagrangian of the system is
\beq {\mathcal L}=\psi^\dagger (i \partial_0 +
\frac{\vec{\nabla}^2}{2 M})\psi -\frac{C_0}{4}(\psi^\dagger\psi)^2
- \frac{D_0}{36}(\psi^\dagger\psi)+\cdots\label{psilagrangian} \eeq where $\psi$
($\psi^\dagger$) is the field that annihilates (creates) a
particle, $C_0$ and $D_0$ are constants that will be determined
later and the dots stand for terms with either more derivatives or $\psi$
fields, whose contributions are suppressed at the order we are considering
here. We will discuss the bosonic case first and comment on the few, but important,
 differences present in the fermionic case later. 
It is convenient to introduce a dummy field $d$ with the
quantum numbers of two particles (a dimer)  and use the equivalent
Lagrangian

\beq {\mathcal L}=\psi^\dagger (i \partial_0 +
\frac{\vec{\nabla}^2}{2 M})\psi+\Delta d^\dagger d
-\frac{g}{\sqrt{2}}(d^\dagger\psi^2 +h.c.)- g_3 d^\dagger
d \psi^\dagger \psi +\cdots\label{dlagrangian}, \eeq
\noindent where $2g^2/\Delta=C_0$ and $36 g_3 g^2/\Delta^2=D_0 $. The
equivalence between the two Lagrangian can  be seen by performing
the gaussian integration over the auxiliary field $d$ and
recovering Eq.~(\ref{psilagrangian}). The value of the constant $\Delta$ is arbitrary
and affects only the normalization of the dimer field. Physical quantities depend on it only
through the combinations $C_0$ and $D_0$.

\begin{figure}[t]
\centerline{\includegraphics*[bbllx=0,bblly=0,bburx=480,bbury=560,scale=0.2,clip=true]
{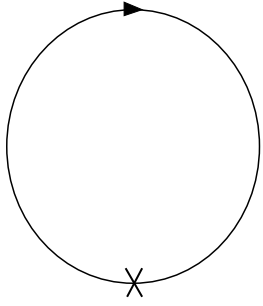}}
\caption{ \label{fign1}
\textit{Graph determining the density at leading order in $z$}.
The cross represents an insertion of the particle number operator. }
\end{figure}

We now compute the particle density $n$ in a expansion in powers of
$z$ and, by comparing with Eq.~(\ref{virialexpansion}), determine
the coefficients $b_l$. The computation of the two first virial
coefficients is rather trivial, and there are fairly explicit
general formulae for them. We will quickly discuss them here in
order to explain the method used in selectingwchich  diagrams contribute at each
order.

In selecting the diagrams contributing to 
given order in $z$ one should  notice that diagrams with a closed particle line 
vanish in vacuum ($z=0$). Consequently, at
low densities their contribution is suppressed by one power of $z$ for each closed particle loop. 
With that observation in mind we see that the only diagram
contributing to the particle density 
at leading order in $z$ is the
loop diagram shown in Fig.~(\ref{fign1}):
\beq n_1&=& T\sum_{k_0} \int
\frac{d^3k}{(2\pi)^3}\frac{-1}{-i k_0 + \epsilon_k}\nonumber\\
&\approx&\frac{z}{\lambdabar^3}+\frac{z^2}{2^{\frac{3}{2}}\lambdabar^3}
+\frac{z^3}{3^{\frac{3}{2}}\lambdabar^3}+{\mathcal O}(z^4),\label{n1}
\eeq 
 The sum over the frequencies $k_0$ is over all integer multiples of $2\pi T$ 
 and $\epsilon_k=\vec{k}^2/2M - \mu$.
Eq.~(\ref{n1}) determines $b_1=1$, which is the free gas result. It also
gives some contributions to $b_2$ and $b_3$.

 The $\mathcal{O}(z^2)$  contributions can be divided into the one-body
 contribution $n_2^{(1)}$ 
 coming from the second term in Eq.~(\ref{n1}) and the
 two-body contributions $n_2^{(2)}$ .
From Eq.~(\ref{n1}) and Eq.~(\ref{virialexpansion}) we find $b_2^{(1)}=2^{-\frac{5}{2}}$.
The diagrams contributing to $n_2^{(2)}$ are shown in Fig.~(\ref{fign2}).
The need for the resummation in the full dimer propagator 
indicated at the bottom of  Fig.~(\ref{fign2})
is more easily explained 
 after computing it. The dimer propagator is given by a geometrical sum
\begin{eqnarray}
{\mathcal D}(p)&=&-\frac{1}{\Delta}+\frac{1}{\Delta}\Sigma(p)\frac{1}{\Delta}+\cdots\nonumber\\
   &=&-\frac{1}{\Delta+\Sigma(p)},
\end{eqnarray}
where
\begin{eqnarray}
\Sigma(p)&=&\Sigma^{(0)}(p)+\Sigma^{(1)}(p)+{\mathcal O}(z^2)\\
&\!\!\!\!\!\!\!\!\!\!\!\!\!= &\!\!\!\!\!\!\!\!\!
\frac{Mg^2}{4\pi}\!\left(\!\frac{2\Lambda}{\pi}\!-\!\sqrt{\frac{\vec{p}^2}{4}\!
-\!2M\!\mu\!-\!iMp_0} 
 \right)\!\! - \!\!\frac{zMg^2}{2\pi^2 p}\int_0^\infty\!\!\!\!\! dk^2  e^{-\frac{\beta k^2}{2M}} 
{\rm arctgh}(\frac{p \ k}{k^2\!+\!\frac{p^2}{2}\!-\!2M\!\mu\!-\!iM\!p_0})+{\mathcal O}(z^2)\nonumber. 
\end{eqnarray}
In particular, the leading order dimer propagator is the same as in the
vacuum
\begin{equation}
{\mathcal D^{(0)}}(p)=\frac{4\pi}{Mg^2}\frac{1}{-\frac{4\pi}{M
g^2}-\frac{2}{\pi}\Lambda+\sqrt{\frac{\vec{p}^2}{4}-2M\mu-iMp_0}}. \label{dpropagator}
\end{equation}
\noindent 
 The particle-particle scattering amplitude in the vacuum
is given by $g^2{\mathcal D}$. Thus, by setting
$4\pi\Delta/Mg^2+2\Lambda/\pi=1/a$, we obtain  the
leading order scattering amplitude in the effective range
expansion (after analytically continuing
the amplitude to real values of the energy by setting 
$k_0\rightarrow k_0-2M\mu+i\epsilon$). Notice the fine tuning
between the linearly divergent term $2\Lambda/\pi$ (kinetic energy) and the
interaction term $4\pi\Delta/Mg(\Lambda)^2$ (potential energy) needed to produce
a small value of $1/a$.
\begin{figure}[t]
\centerline{\includegraphics*[bbllx=0,bblly=0,bburx=626,bbury=425,scale=0.35,clip=true]
{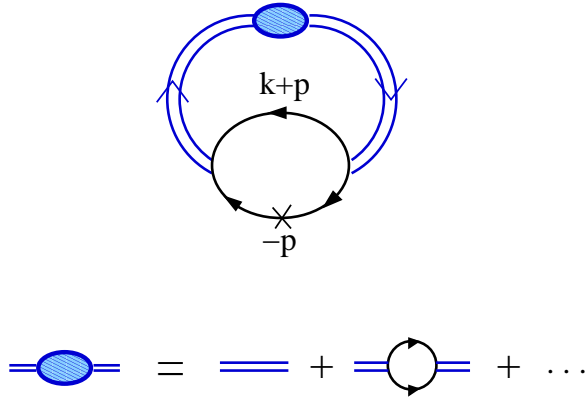}}
\caption{ \label{fign2}
\textit{Diagram determining the density at leading order in $z^2$} (top)
and the dressing of the dimer propagator. }
\end{figure}
We can understand now the reason for the resummation of graphs at the 
bottom of  Fig.~(\ref{fign2}).
For momenta of order $p\sim \sqrt{MT}\sim 1/a$, the square root in the denominator of
Eq.~(\ref{dpropagator}), generated by the loops in the dimer propagator, 
 is {\it not} negligible compared to $1/a$.

 Plugging  the expression in
Eq.~(\ref{dpropagator}) in the diagram in Fig.~(\ref{fign2}) we have
\begin{eqnarray}
n_2^{(2)}&=& M T \sum_{k_0}\int \frac{d^3k}{(2\pi)^3}
\frac{1}{-\frac{1}{a}+\sqrt{\frac{\vec{k}^2}{4}-2M\mu-iMk_0}}
\frac{1}{\sqrt{\frac{\vec{k}^2}{4}-2M\mu-iMk_0}}\nonumber\\
&=&M T \int \frac{d\eta}{2\pi i}\int \frac{d^3k}{(2\pi)^3}
\frac{1}{e^{\beta\eta}- 1}
\frac{1}{-\frac{1}{a}+\sqrt{\frac{\vec{k}^2}{4}-2M\mu-M\eta}}
\frac{1}{\sqrt{\frac{\vec{k}^2}{4}-2M\mu-M\eta}}\ .
\label{n2_intermediate}
\end{eqnarray}
The integration over $\eta$, on a contour encircling the imaginary
axis, can be deformed  into an integral over the discontinuity on
the branch cut (which describes the dimer break up) plus the
contribution of the dimer pole. The final result is 
\begin{equation}
 n_2^{(2)}=2 b_2 \frac{z^2}{\lambdabar^3}  
=\frac{z^2}{\lambdabar^3}\ \   e^{\beta B_2}\ \  2^{3/2}  
\left(1+ \mathrm{ Erf}\left(\frac{1}{a\sqrt{MT}}\right)\right)\label{n2}\ , \end{equation} 
where $B_2=1/(M a^2)$ is the location of the pole in the dimer propagator.
This pole corresponds to a bound state  (if $a$ is positive) or
a ``virtual'' bound state, that is, a pole in the unphysical
sheet (if $a$ is negative).
Eq.~(\ref{n2}) is a particular case of the classic formula
relating $b_2^{(2)}$ to the two-particle scattering phase shift 
~\cite{beth_uhlenbeck_virial} 
\beq
b_2^{(2)}= 2^{1/2}\left(e^{\beta B_2}+\frac{1}{\pi}\int_0^\infty dk
\frac{d \delta(k)}{dk} e^{-\frac{\beta k^2}{M}}\right)\ , 
\eeq
in the case where the phase shifts are given by the
leading order effective range expansion $k \cot \delta(k)=-1/a$. A
few  points are worth mentioning here. First, $b_2$ is a
continuous function of $a$ at $1/a=0$. The change in the potential needed
to take $a$ from a large and positive value to a large and negative value is
small, and that change has only a small effect in the thermodynamics,
even though there is a qualitative difference in the spectrum
(from a bound state to a virtual bound state). Second, $b_2$ is
positive and, for $B_2>T$, it is exponentially large. 
In the case $a>0$, this is easily understood as a
consequence of the existence of a bound state: when $T\ll B_2$ most 
particles form tight two-body bound states therefore increasing
the density for a fixed temperature and chemical potential. In
fact, if the $z^2$ terms dominate, the ratio of the two equations
in Eq.~(\ref{virialexpansion}) give the equation of state of a
free gas with density $n/2$.

  The third virial coefficient $b_3$  includes three pieces. The first, $b_3^{(1)}$, 
is determined by the $z^3$ piece in Eq.~(\ref{n1}) and equals $b_3^{(1)}=3^{-\frac{5}{2}}$.
The second one, $b_3^{(2)}$,   comes from a subleading piece of the diagram on  Fig.~(\ref{fign2})
and is given by
\begin{eqnarray}\label{b_3^2}
n_3^{(2)}&=& 3\ b_3^{(2)}\ \frac{z^3}{\lambdabar^3}\\
&=&2\ T\sum_{k_0} \int\frac{d^3k}{(2\pi)^3} 
\left[({\mathcal D}^{(0)}(k))^2 \Sigma^{(1)}(k)\frac{d}{idk_0}\Sigma^{(0)}(k) 
+{\mathcal D}^{(0)}(k)\frac{d}{idk_0}\Sigma^{(1)}(k)  \right]  \nonumber\\
&=& 2\ T\sum_{k_0} \int\frac{d^3k}{(2\pi)^3} 
\frac{d}{dk_0}( {\mathcal D}^{(0)}(k)\Sigma^{(1)}(k))\nonumber\\
&=&\frac{i\beta z^3}{\pi^4}\int_0^\infty dk^2\int_0^\infty dp\int d\zeta e^{-\beta\frac{p^2}{2M}}
{\rm arctgh}(\frac{p \ k}{k^2+\frac{p^2}{2}-2M\mu-iMp_0})
\frac{1}{-\frac{1}{a}+\sqrt{\frac{k}{4}-M\zeta}}\nonumber,
\end{eqnarray}
where the integral over $\zeta$ runs on a vertical line just to the left of the pole at $-B_2$.
The integral in Eq.~(\ref{b_3^2}) cannot be expressed analytically and has to be computed numerically.
\begin{figure}[t]
\centerline{\includegraphics*[bbllx=0,bblly=0,bburx=826,bbury=536,scale=.35,clip=true]
{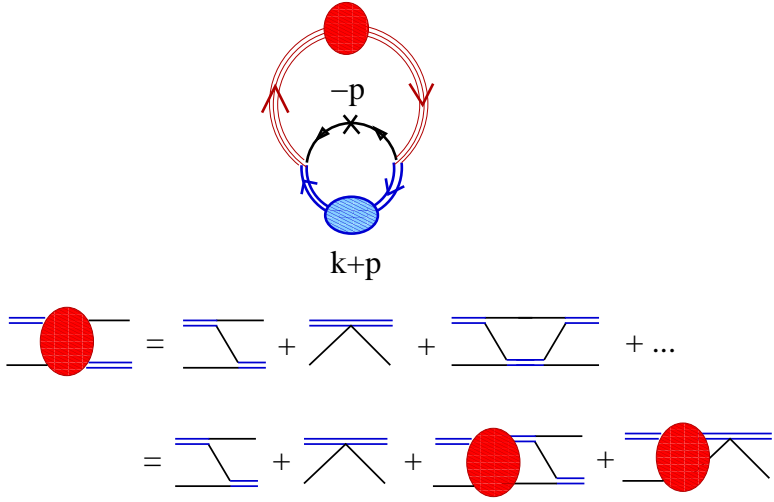}}
\caption{ \label{fign3}
\textit{Graph determining the density at leading order in $z^3$ and the integral
equation for the three-body amplitude.} }
\end{figure}
The third piece, $b_3^{(3)}$, comes from actual three-body correlations in the system, and brings in
new physics besides that contained in two-body scattering. The diagrams contributing to it
are shown on Fig.~(\ref{fign3}).
Just as the two-body scattering amplitude
cannot be computed in perturbation theory, the three-body
 amplitude that enters in Fig.~(\ref{fign3}) also involves a resummation
of an infinite number of diagrams.
 This is because each additional loop involves a factor of $p a$, 
where $p$ is the typical momentum flowing through the loop. In
the case considered here $Q\sim 1/\lambdabar\sim 1/a$ 
and the ``suppression'' factor is of order $1$. 
The diagrams in Fig.~(\ref{fign3}) gives
\begin{eqnarray}
n_3^{(3)}&=& T^2 \sum_{k_0,p_0} \int\!\! \frac{d^3k}{(2\pi)^3}\frac{d^3p}{(2\pi)^3}
\left(\frac{1}{ip_0+\epsilon_p}\right)^2 {\mathcal D}^{(0)}(p+k){\mathcal T}(-p,p+k;-p,p+k)\\
&=&T^2 \sum_{k_0} \int\!\! \frac{d^3k}{(2\pi)^3}\frac{d^3p}{(2\pi)^3}\frac{d\eta}{2\pi i}
\frac{1}{e^{\beta\eta}-1}
\left(\frac{1}{\eta+\epsilon_p}\right)^2 {\mathcal D}^{(0)}(p+k){\mathcal T}(-p,p+k;-p,p+k)\nonumber
\end{eqnarray}
where  in the second line $p=(-i\eta,\vec{p})$ and ${\mathcal T}(-p,p+k;-p,p+k) $ is the 
forward particle-dimer scattering amplitude determined by the diagrams
 at the bottom of Fig.~(\ref{fign3}).
The integral over $\eta$ is dominated by the particle pole at $\eta=-\epsilon_p$,
with the contribution coming  from the dimer pole and cut
suppressed by two powers of $z$. We then have
\begin{eqnarray}\label{n3graph}
n_3^{(3)}&=&-\frac{2\pi z^3}{g^2}\int\frac{d^3k}{(2\pi)^3}\frac{d^3P}{(2\pi)^3}
\int\frac{d\zeta}{2\pi i}
e^{-\beta\zeta}\ {\mathcal T}(-P+\frac{k}{3},P+\frac{2k}{3};-P+\frac{k}{3},P+\frac{2k}{3})
\vert_{k_0=i3\mu-i\zeta} \nonumber\\
& &\ \ \ \ \ \ \ \ \ \ \frac{1}{-\frac{1}{a}+\sqrt{\frac{3 \vec{P}^2}{4}+\frac{\vec{k}^2}{6}-M\zeta}}  
\frac{1}{\sqrt{\frac{3 \vec{P}^2}{4}+\frac{\vec{k}^2}{6}-M\zeta}}  ,
\end{eqnarray}
where we use the center-of-mass variable $P=(i\epsilon_p,\vec{p}+\vec{k}/3)$. The integration
over the new variable $\zeta=3\mu+\zeta$ in on a vertical line just to the left of the
left-most singularity that can be a trimer pole at $k^2/6M-B_3$ or a dimer pole 
at $k^2/6M+3P^2/4M-B_2$. The cuts in the three-body amplitude describing dimer-particle scattering, 
dimer breakup, etc., are on the real axis, to the right of $2$ and $3$ particle poles.

Unlike the dimer propagator, the diagrams adding up to the three-body amplitude 
${\mathcal T}(-p,p+k;-p,p+k)$ do not form a simple geometrical series and cannot be summed 
up analytically.
However, their sum is  determined by the integral equation depicted at the bottom of Fig.~(\ref{fign3}).
This integral equation  (Faddeev equation)
is particularly simple for the separable potential used here \footnote{Some simplifications
in the computation of $b_3$ in the case of separable potentials were noted in ~\cite{reiner}} 
and was derived many times before
with a variety of techniques 
~\cite{skorny,bedaque_bira_ndI,bedaque_bira_review}. 
It is more easily written in terms of the $s-$wave amplitude ${\mathcal T}_0(\zeta,P,P')$ defined by
\begin{equation}\label{T0}
Mg^2\frac{{\mathcal T}_0(\zeta,P,P')}{\vec{P}^2-\frac{4M}{3}(\zeta+B_2)}
=\int \frac{d\hat{P}}{4\pi}\frac{d\hat{P'}}{4\pi} 
\frac{{\mathcal T}(-P+\frac{k}{3},P+\frac{2k}{3};-P+\frac{k}{3},P+\frac{2k}{3})
\vert_{k_0=i3\mu-i\zeta}}
{-\frac{1}{a}+\sqrt{\frac{3 \vec{{P'}}^2}{4}+\frac{\vec{k}^2}{6}-M\zeta}},
\end{equation} Higher partial 
waves, describing more peripheral particle-dimer collisions,  give  suppressed contributions.
${\mathcal T}_0(\zeta,P,P')$ satisfies
\begin{equation}\label{int_equation}
{\mathcal T}_0(\zeta,P,P')=K(\zeta,P,P') +\frac{2}{\pi}\int_0^\Lambda dqq^2 
\frac{{\mathcal T}_0(\zeta,P,q)}{q^2-\frac{4M}{3}(\zeta+B_2)}  K(\zeta,q,P'),
\end{equation}with the kernel
\begin{equation}\label{kernel}
K(\zeta,P,P')=\frac{4}{3}\left( \frac{1}{a}+\sqrt{\frac{3P^2}{4}-M\zeta  } \right)
\left[ \frac{1}{PP'}\log \left( \frac{P^2+PP'+{P'}^2-M\zeta}
                                     {P^2-PP'+{P'}^2-M\zeta} \right)
                          -\frac{g_3(\Lambda)}{Mg^2} 
\right]
\end{equation} 
Eq.~(\ref{int_equation}) has a number of surprising properties 
~\cite{faddeev_minlos,danilov,efimovI,efimovII,amado_noble}. 
This properties are more naturally described in terms of renormalization theory.
It was shown in ~\cite{3stooges_bosons_short, 3stooges_bosons_long} that 
${\mathcal T}_0(\zeta,P,P')$ is kept cutoff independent
for small values of $P,P'\ll \Lambda$ if, and only if, the three-body force varies with the cutoff
$\Lambda$ as 
\begin{equation}\label{RG}
-\frac{g_3(\Lambda)\Lambda^2}{2Mg^2}=\frac{\sin(s_0 \log(\Lambda/\Lambda^*)-\arctan s_0)}
                                       {\sin(s_0 \log(\Lambda/\Lambda^*)+\arctan s_0)}
+{\mathcal O}(\frac{1}{a\Lambda}),
\end{equation} where $\Lambda^*$ is a new parameter that cannot be measured through two-particle
experiments and $s_0\approx 1.006$ is the solution of a certain trigonometric equation.
The failure to include the three-body force term leads to a strong cutoff dependence on the results
and thus, to the impossibility of arriving at model independent results.
The three-body scattering amplitude depends, even at the leading order in the expansion in
powers of $\lambdabar/R$, on the value
of the three-body force, here parameterized by $\Lambda^*$. 
The parameter  $\Lambda^*$ encodes all short distance physics, besides the value of $a$,
 necessary to describe three-body systems. Like $a$, it varies from one system to another
and only the measurement of some three-body observable allows us to fix it.
The appearance of an extra parameter not determined by
two-body scattering precludes the possibility of predictions in the three-body sector
based only on the value of $a$.

 Using Eq.~(\ref{n3graph}) and Eq.~(\ref{T0}) and performing a trivial integral over $k$ and
the angular part of $\vec{P}$ we arrive at
\begin{equation}\label{n_3^3}
n_3^{(3)}=3 b_3^{(3)} \frac{z^3}{\lambdabar^3}
=i\frac{9\sqrt{3}}{4\pi^2}M\frac{z^3}{\lambdabar^3}
\int_0^\infty \!\!dP P^2\!\!\int \!\!d\zeta e^{\beta\zeta} \frac{{\mathcal T}_0(\zeta,P,P)}
                                                    {(\frac{3P^2}{4}-M\zeta-B_2)^2}
\left( 1+\frac{1}{a}\frac{1}{\sqrt{\frac{3P^2}{4}-M\zeta}} \right)\nonumber
\end{equation} 
The amplitude ${\mathcal T}_0(\zeta,P,P) $ and the quadratures in Eq.~(\ref{n_3^3}) have to be
performed numerically. This is not computationally demanding and a Mathematica notebook that 
performs this task can be downloaded from {\tt http://www-nsdth.lbl.gov/\~bedaque}.

The coefficient $b_3=b_3^{(1)}+b_3^{(2)}+b_3^{(3)}$ is a function of the temperature and 
of the parameters describing
the microscopic dynamics (the scattering length $a$ and the three-body parameter $\Lambda^*$).
In Fig.~(\ref{figb3}) we show the value of $b_3$ as a function of $1/a$ 
for a few values of these parameters.
As an example we pick $M=M_{^4{\rm He}}$, $T=10^{-7}{\rm eV}\approx 1.16 mK$ and 
$-\Lambda^2 g_3(\Lambda)/(2 M g^2)\equiv H(\Lambda)=-3.235$ (lower blue solid curve), 
$-0.22$ (intermediate red solid curve) and $0.05$ (upper black solid curve). In all of them
we use $\Lambda=200$ eV, but, as mentioned above, the results are cutoff independent up to 
small corrections of  order  $1/\Lambda a$. The first of these values was chosen following 
~\cite{braaten_hans_4he}
and it gives rise to a trimer with the binding energy of the shallower $^4$He trimer (as predicted
through potential model calculations ~\cite{roudnev,motovilov_4he,lewerenz} ) 
when $a$ is set to the value inferred from the 
dimer measurement $a=10^{+8}_{-18}\mathring{A}$  ~\cite{grisenti_4he_dimer}. 
For the value of the $\Lambda$ used the deeper trimer state is absent.

When $a$ and $\Lambda^*$ are such that there 
is a three-body bound state $B_3\agt B_2$, 
we find empirically that $b_3$ scales
roughly as $e^{\beta B_3}$, see Fig.~\ref{figb3}. This is not surprising and is a direct analogue  
of the physics
described by Eq.~(\ref{n2}) in the case of the coefficient $b_2$: when 
$\beta B_3 \gg 1$
most particles form three-body bound states, which
 increases the density for a given
temperature and chemical potential as compared to the free gas. 
This does {\it not} mean, however, that the integral in Eq.~(\ref{n_3^3}) is dominated by the trimer
pole: the contribution from the two-and three-particle cuts is never negligible.
For values of $a \Lambda^*$ for which $B_3$ approaches $B_2$ the dimer-particle cross section 
diverges. Still, the 
three-particle correlations described by $b_3$ are smooth at those points.
  For systems with a trimer deeper than the dimer, $b_3$ can be much larger that both
$b_1$ and $b_2$, and dominate the virial expansion. This is
not a rare situation, due to a well known but strange feature of the three-resonating particle 
system: as the potential is changed to make the dimer {\it shallower} , the trimer becomes
{\it deeper} ~\cite{efimovI}. This is what happens in the $H=0.05$ case shown in the 
upper (black) solid curve in Fig.~\ref{figb3}.

 In atomic traps close to a Feshbach resonance it 
is $a$ that is a tunable parameter. The influence of the magnetic field on the value of the
three-body force parameter $\Lambda^*$ is small, since that describes short-distance physics
of an energy scale much higher than the magnetic field, and can be disregarded in a 
first approximation. Thus we can regard the horizontal axis in Fig.~(\ref{figb3}) as denoting
the magnetic field using the Feshbach resonance formula
\beq
a(B)=a(B=0)\left(1+\frac{\delta B}{B-B_0}\right).
\eeq

\begin{figure}[t]
\centerline{\includegraphics*[bbllx=69,bblly=475,bburx=472,bbury=721,scale=.75,clip=true]
{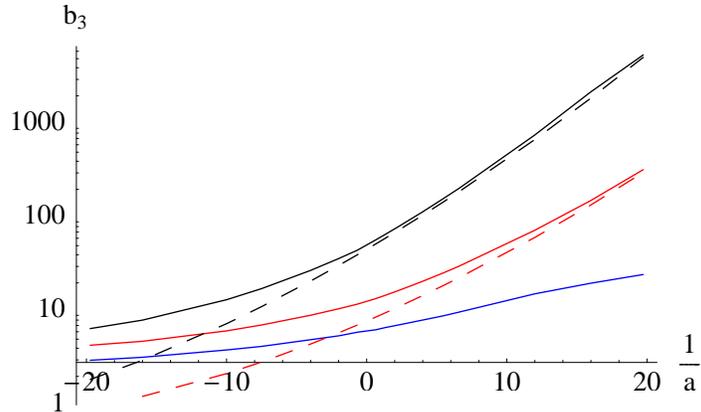}}
\caption{ \label{figb3}
\textit{$b_3$ as a function of the inverse scattering length for three values
of $H(\Lambda=200{\rm eV})$ (solid lines, $H=-3.23, -0.22, 0.05$, from bottom to top).
The dashed line is the estimate $e^{\beta B_3}$.}}
\end{figure}

Let us summarize the approximation performed here that allowed such a simple evaluation of
such a complex three-body dynamics. First, we approximate the two-body potential by a
short range, delta-function-like interaction. Corrections to this approximation are 
proportional to the effective range $ r_0/\lambdabar$, where  $r_0\sim R$ is the effective range, 
much smaller than $a$ by assumption.
Effective range corrections can be easily included, even to very high orders,
 without spoiling the simplicity of 
the method, as it has been done in few-nucleon physics ~\cite{bedaque_bira_review,seattle_pionless}. 
The second approximation was the 
non-inclusion of higher partial waves. In the two-body sector, $p$-waves and higher are 
suppressed by at lest two powers of $ R/\lambdabar$,  and are small. 
In the three-body sector the suppression
of the particle-dimer $l$-wave interactions are not suppressed by powers of $R/\lambdabar$, but
by $1/(l+1)$. 
All the higher partial waves correspond to a repulsive kernel in Eq.~(\ref{int_equation}),
which do not support  bound states and cannot produce an enhancement of the form 
$e^{\beta B_3}$. In practice the phase shifts are rather small but, if needed, they
 can also be easily included by solving the analogue of 
Eq.~(\ref{int_equation}) 
corresponding to the higher partial wave and adding it to Eq.~(\ref{n3graph}).
 
Our results are valid only in 
true thermal equilibrium, after two- and three-particle bound states had the time to form, 
and assuming that they stay in the system ( do not escape from the trap, if that is the case). 
For systems like the alkali atoms
studied in magnetic/optical traps,  that have deep bound states with typical interparticle distance
of order $\sim R$ that lie
 outside the validity range of our effective theory, this means that our results
are relevant only for the metastable state before the collapse of the system, but after 
the two- and three-body bound states form. The formation of a $l-$body bound state
requires the approach of $l+1$ particles in order to conserve energy and momentum, 
and their rates are  consequently suppressed by $n^{l+1}$. These rates are not known at 
finite temperature but have been studied at zero temperature in 
~\cite{bedaque_recombination,moerdijk,fedichev_recombination} (recombination into shallow states) 
and ~\cite{braaten_hans_recombination_deep} (recombination into deep states). In cases where
$a>0$ the recombination rate into deep (two-body) bound states is estimated to be much smaller
than the rate into shallow bound states suggesting that there is a time window in which
our results apply. In the $a<0$ case there is no two-body bound state and it is not known how 
the rates for the formation of deep and shallow three-body bound states compare.

Finally, let us consider the changes introduced in case of fermionic particles. 
In the non-degenerate case considered here, the effect of the quantum statistics
in the thermodynamics is minor, and amounts to 
 a change in sign in some of the coefficients $b_{1,2}^{(1,3,...)}$. 
The elementary collisions, however,
may differ a great deal due to the exclusion principle. If we have only one fermionic
species in the system, $s$-wave scattering is impossible and all virial coefficients 
$b_2,b_3,\cdots$ are suppressed. In the case of two fermionic species 
(as a dilute neutron gas), two-body
collisions are possible and the standard result in Eq.~(\ref{n2}) is valid. The physics of
the three-body correlations is however, very different. No three-body force term without 
derivatives exist, and the three-body force contribution is suppressed. $b_3$ can be computed
in terms of $a$ alone, but cannot ever be large and dominate the expansion, as the
kernel appearing in Eq.~(\ref{int_equation}) would be repulsive and would not support a bound state.
The case with three or more fermionic species  with all the scattering lengths 
large but not necessarily equal, which includes the
dilute nuclear matter case (protons and neutrons with spin either up or down), 
is very similar to the bosonic case. 
 Three particles can occupy the same point in space without violating the exclusion principle
and, as a consequence, the two coupled equations that substitute Eq.~(\ref{int_equation}) have very 
similar properties to the bosonic equation ~\cite{3stooges_triton}.

\begin{acknowledgments}
This work was supported by the Director, Office of Energy Research, 
Office of High Energy and Nuclear Physics, and by the Office of 
Basic Energy Sciences, Division of Nuclear Sciences, 
of the U.S. Department of Energy under Contract No.
DE-AC03-76SF00098. 
\end{acknowledgments}

\end{section}

\bibliographystyle{apsrev}
\bibliography{virial}

\end{document}